\documentclass[useAMS,usenatbib]{mn2e}
\usepackage[dvips]{epsfig}
\usepackage[]{natbib}
\usepackage[]{fixltx2e}
\usepackage{amsmath}
\usepackage{multirow}
\title[Gas inflows towards the nucleus of NGC\,1358]{Gas inflows towards the nucleus of NGC\,1358}
\author[A. Schnorr-M\"uller et al.]
  {Allan Schnorr-M\"uller$^1$,  Thaisa Storchi-Bergmann$^1$, Neil M. Nagar$^2$, Andrew Robinson$^3$,  
  \newauthor and Davide Lena$^{4,5}$\\
  $^1$Instituto de F\'isica, Universidade Federal do Rio Grande do Sul, 91501-970, Porto Alegre, RS, Brazil\\
  $^2$Astronomy Department, Universidad de Concepci\'on, Casilla 160-C, Concepci\'on, Chile\\
  $^3$Physics Department, Rochester Institute of Technology, Rochester, New York 14623, USA\\
  $^4$SRON Netherlands Institute for Space Research, Sorbonnelaan 2, NL-3584 CA Utrecht, the Netherlands \\
  $^5$Department of Astrophysics/IMAPP, Radboud University, Nijmegen, PO Box 9010, NL-6500 GL Nijmegen, the Netherlands \\}

\pagerange{\pageref{firstpage}--\pageref{lastpage}} \pubyear{2012}

\begin{document}

\label{firstpage}
\maketitle

\begin{abstract}
We use optical spectra from the inner 1.8\,$\times$\,2.5\,kpc$^2$ of the Seyfert\,2 galaxy NGC\,1358, obtained with the GMOS integral field spectrograph on the Gemini South telescope at a spatial resolution of $\approx$\,165\,pc, to assess the feeding and feedback processes in this nearby active galaxy. Five gaseous kinematical components are observed in the emission line profiles. One of the components is present in the entire field-of-view and we interpret it as due to gas rotating in the disk of the galaxy. Three of the remaining components we interpret as associated to active galactic nucleus (AGN) feedback: a compact unresolved outflow in the inner 1\arcsec\ and two gas clouds observed at opposite sides of the nucleus, which we propose have been ejected in a previous AGN burst. The disk component velocity field is strongly disturbed by a large scale bar. The subtraction of a velocity model combining both rotation and bar flows reveals three kinematic nuclear spiral arms: two in inflow and one in outflow. We estimate the mass inflow rate in the inner 180\,pc obtaining $\dot{M}_{in}$\,$\approx$\,1.5\,$\times\,10^{-2}$M$_{\odot}$\,yr$^{-1}$, about 160 times larger than the accretion rate necessary to power this AGN. 

\end{abstract}


\section{Introduction}

It is widely accepted that the radiation emitted by an active galactic nucleus (AGN) is a result of accretion onto the central supermassive black hole (hereafter SMBH). However, the mechanisms involved in the transfer of mass from kiloparsec scales down to nuclear scales are not well understood. The ubiquity of dust structures (spirals, filaments and disks) in the inner kiloparsec of AGN suggest these structures are likely associated shocks and angular momentum dissipation in the interstellar medium, thus tracing the transfer of gas to the inner tens or hundreds parsecs \citep{martini03,lopes07}. This is supported by simulations, which showed that, if a central SMBH is present, spiral shocks can extend all the way to the SMBH vicinity and generate gas inflow consistent with the observed accretion rates \citep{maciejewski04a,maciejewski04b}. 

In order to assess the role of nuclear dust structures, in particular nuclear spirals, in the transport of gas to the AGN, our group has been mapping gas flows in the inner kiloparsec of nearby AGN using optical and near-infrared integral field spectroscopic observations. So far, we have observed gas inflows along nuclear spirals in NGC\,1097 \citep{fathi06}, NGC\,6951 \citep{thaisa07}, NGC\,4051 \citep{rogemar08}, M\,79 \citep{rogemar13}, NGC\,2110, \citep{allan14a}, NGC\,7213 \citep{allan14b} and NGC\,1667 \citep{allan17}. We have also observed gas inflows in the galaxy M\,81 \citep{allan11}, where the inflow was mostly traced by dust lanes and in NGC\,3081 \citep{allan16}, where a nuclear bar is feeding the AGN. There is tentative evidence of inflows in NGC\,1386 \citep{lena15}. Gas inflows have also been observed by other groups. Near-infrared integral field spectroscopic observations revealed inflows along nuclear spiral arms in NGC\,1097 \citep{davies09}, NGC\,5643 \citep{davies14} and NGC\,7743 \citep{davies14}, and gas inflow along a bar in NGC\,3227 \citep{davies14}. Recent ALMA observations of molecular gas revealed streaming motions along nuclear spirals in NGC\,1433 \citep{combes13} and NGC\,1566 \citep{combes14}. Observations of CO also revealed gas inflows in NGC\,1068 \citep{burillo14}, NGC\,2782 \citep{hunt08}, NGC\,3147 \citep{casasola08}, NGC\,3627 \citep{casasola11}, NGC\,4579 \citep{burillo09} and NGC\,6574 \citep{krieg08}.

In the present work, we report results obtained from optical integral field spectroscopic observations of the nuclear region of NGC\,1358, a barred S0a galaxy harboring a Seyfert 2 AGN. Observational properties of NGC\,1358 are listed in Table\,1.  

The present paper is organized as follows. In Section \ref{Observations} we describe the observations and data reduction. In Section \ref{Results} we present the procedures used for the analysis of the data and the subsequent results. In section \ref{Discussion} we discuss our results and present estimates of the mass inflow rate and in Section \ref{Conclusion} we present our conclusions.

\begin{figure*}
\includegraphics[scale=0.8]{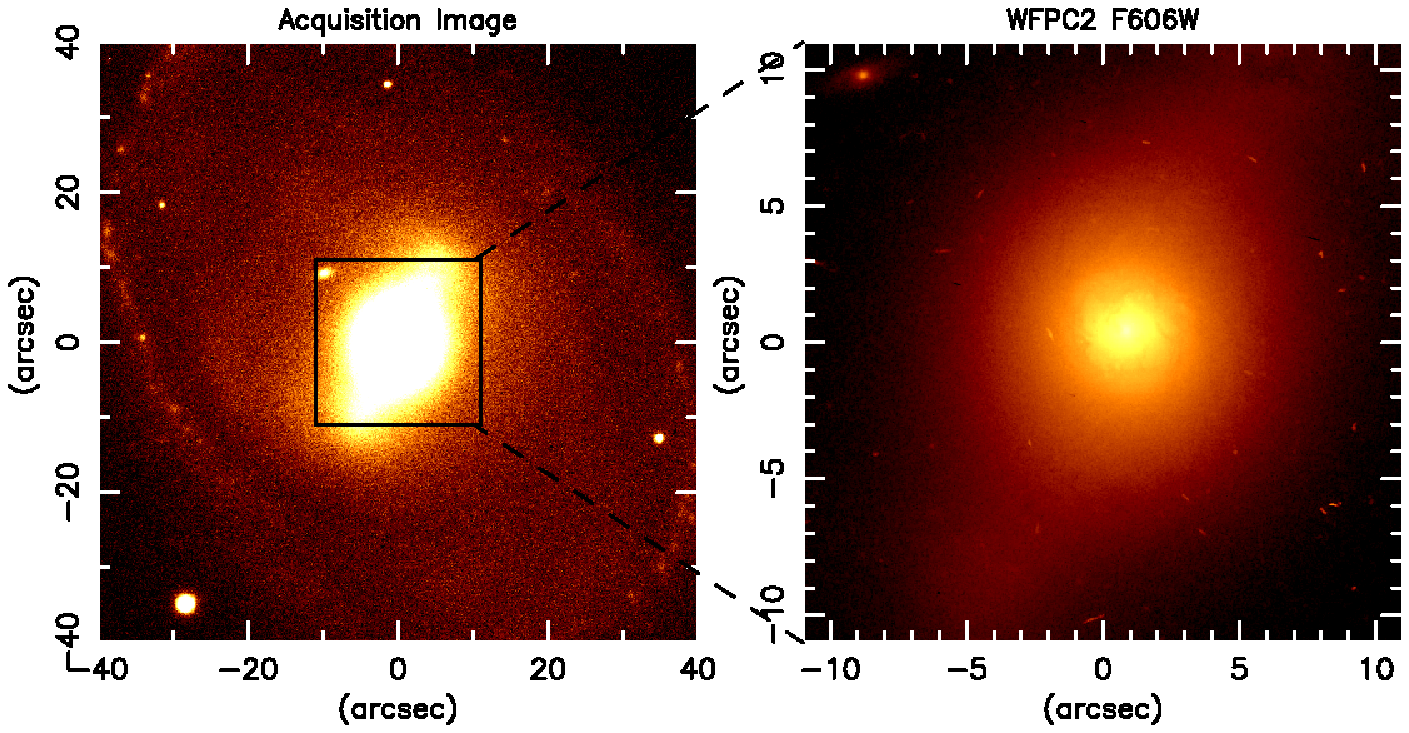}
\includegraphics[scale=0.8]{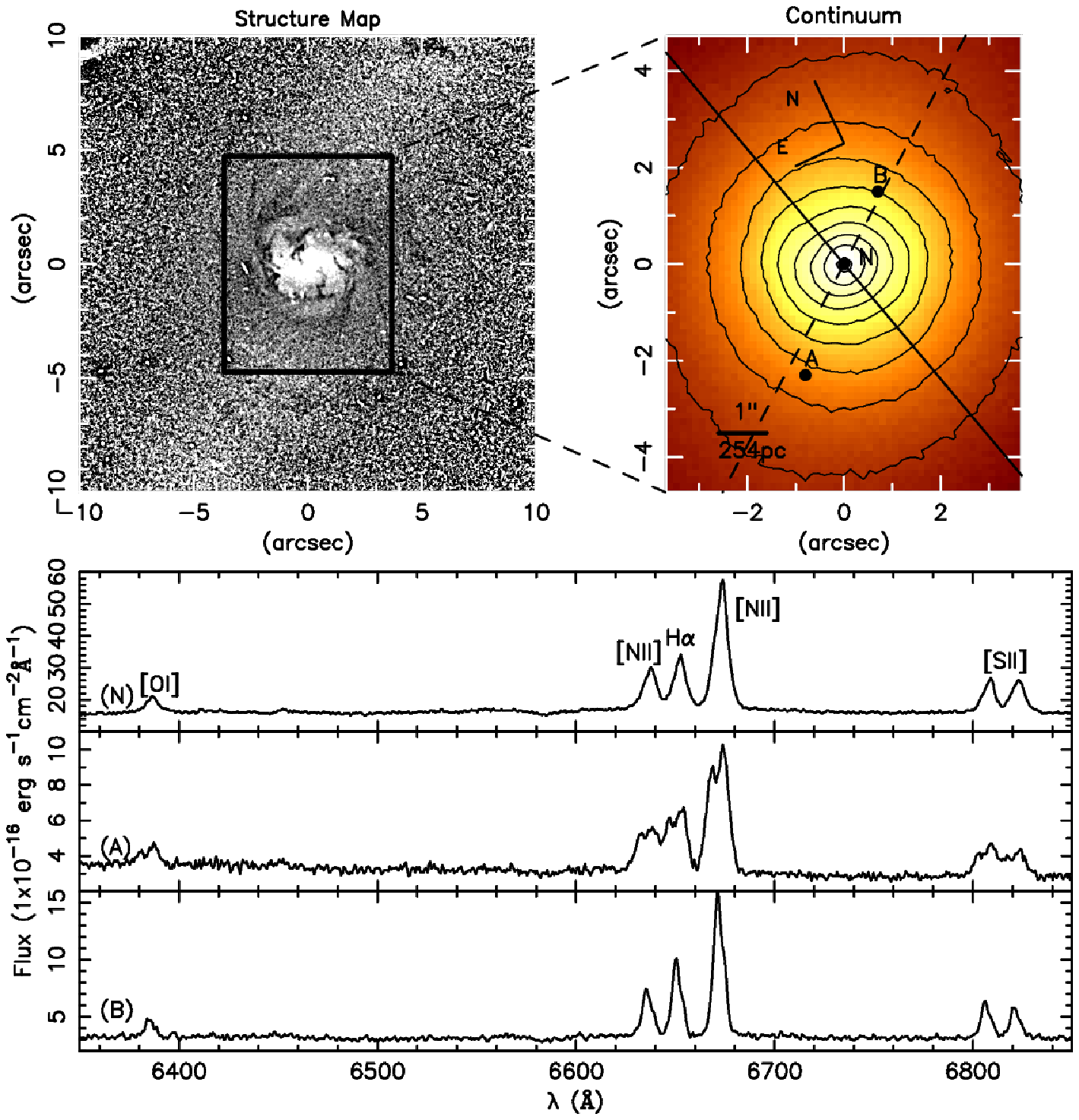}
\caption[Large scale image of NGC1358]{Top left: acquisition image of NGC1358. Top right: WFPC2 image. Middle left: structure map. The rectangle shows the field of the IFU observation. Middle right: continuum image from the IFU spectra (flux in erg\,cm$^{-2}$\,s$^{-1}$ per pixel). The straight black line indicates the position angle of the photometric major axis of the galaxy (PA\,=\,15\ensuremath{^\circ}, see Table\,1) and the dashed black line indicates the position of the bar (PA\,=\,135\ensuremath{^\circ}). Bottom: spectra corresponding to the regions marked as N, A and B in the IFU image. All images have the same orientation, where north makes an angle of 25\ensuremath{^\circ} with the vertical axis.}
\label{fig1}
\end{figure*}

\section {Observations and Data Reduction}\label{Observations}

The observations were obtained with the Integral Field Unit of the Gemini Multi Object Spectrograph (GMOS-IFU) at the Gemini South telescope on the night of January 27, 2011 (Gemini project GS-2010B-Q-19, P.I. Neil M. Nagar), in two-slit mode. The observations consisted of two adjacent IFU fields (covering 7\,$\times$\,5\,arcsec$^{2}$ each) totaling an angular coverage of 7\,$\times$\,10\,arcsec$^{2}$. Six exposures of 350 seconds were obtained. The spectral coverage is 5600-7000\,\r{A} at a resolving power of R\,$\approx$\,2000. The seeing during the observation was 0\farcs65, as measured from the FWHM of a spatial profile of the calibration standard star. This corresponds to a spatial resolution at the galaxy of 165\,pc. 

The data reduction was performed using the \textsc{gemini.gmos} package in \textsc{iraf}\footnote{\textit{IRAF} is distributed by the National Optical Astronomy Observatories, which are operated by the Association of Universities for Research in Astronomy, Inc., under cooperative agreement with the National Science Foundation.}. This package is provided by the Gemini Observatory and it is specifically developed for data reduction of observations taken with the GMOS instrument. The data reduction process comprised bias and sky subtraction, flat-fielding, trimming, wavelength and flux calibration and building and combination of the data cubes. The final datacube has a spatial sampling of 0\farcs1\,$\times$\,0\farcs1, containing 7030 spectra. 

\section{Results}\label{Results}

\begin{center}
\begin{table}
\caption{Observational properties of NGC\,1358.}
\tabcolsep=0.085cm
\begin{tabular}{|l|c|c||}
    \hline
    \hline
    RA			     & 03h33m39.7s & \multirow{2}{*}{\citet{argyle90}}\\
    DEC			     & -05d05m22s \\ 
    V$_{sys}$                & $4028\,\pm$10\,km\,s$^{-1}$ & \citet{theureau98}\\
    Morph. Type              & SAB(r)0/a & \citet{rc3}\\
    Activity                 & Sy\,2& \citet{veron06} \\
    Distance (Mpc)           & 53.7 & \citet{theureau98}\\
    Proj. scale (pc/\arcsec) & 254 & NED\footnote{NASA/IPAC extragalactic database}\\
    Inclination              & 54\ensuremath{^\circ}  & \citet{gerssen03}\\ 
    Major axis P.A.          & 15\ensuremath{^\circ} & \citet{gadotti07}\\
    Bar P.A.                 & 135\ensuremath{^\circ} & \citet{gadotti07}\\
    \hline
  \end{tabular}
\end{table} 
\end{center}

In Fig.\,\ref{fig1} we present in the upper left panel the acquisition image of NGC\,1358 and in the upper right panel an image of the inner 22\arcsec\,$\times\,$22\arcsec\ of the galaxy obtained with the Wide Field Planetary Camera 2 (WFPC2) through the filter F606W aboard the Hubble Space Telescope (HST). A large-scale bar is visible in both images, oriented along the position angle (PA) 135\ensuremath{^\circ} (see Table\,1). Faint spiral arms are also visible in the acquisition image. An H$\alpha$ image of NGC\,1358 \citep{delgado97} shows these arms emerge from the bar and are traced by H\,II regions. In the middle left panel we present a structure map of the WFPC2 HST image (see \citealt{pogge02}). The rectangle in this panel shows the field-of-view (hereafter FOV) covered by the IFU observations. Chaotic nuclear spirals arms traced by (dark) dust lanes are a prominent feature in the structure map in the inner 4\arcsec. In the middle right panel we present an image from our IFU observations obtained by integrating the continuum flux within a spectral window from $\lambda$6470\,\r{A} to $\lambda$6580\,\r{A}. The dashed black line traces the orientation of the large-scale bar. The straight black line traces the position of the photometric major axis, oriented along PA\,=\,15\ensuremath{^\circ} (see Table\,1). In the lower panel we present three spectra of the galaxy corresponding to locations marked as A, B and N in the IFU image, showing the complex line profiles of [O\,I]$\lambda$$\lambda$\,6300,6363\,\r{A}, [N\,II]\,$\lambda$$\lambda$6548,6583\,\r{A}, H$\alpha$ and [S\,II]\,$\lambda$$\lambda$6717,6731\,\r{A} observed within the inner 2\arcsec. These spectra were extracted within apertures of 0\farcs3\,$\times\,$0\farcs3. 

The spectrum corresponding to the nucleus (marked as N in Fig.\,\ref{fig1}) is typical of the inner 1\arcsec, where the line profiles have a ``triangular" shape, with a broad base and a narrow top, distinct from a single Gaussian profile. In the spectrum from location A, the line profiles are double peaked. In the spectrum from location B the line profiles are asymmetric, with a ``red shoulder".

\subsection{Measurements}

\begin{figure*}
\includegraphics[scale=0.7]{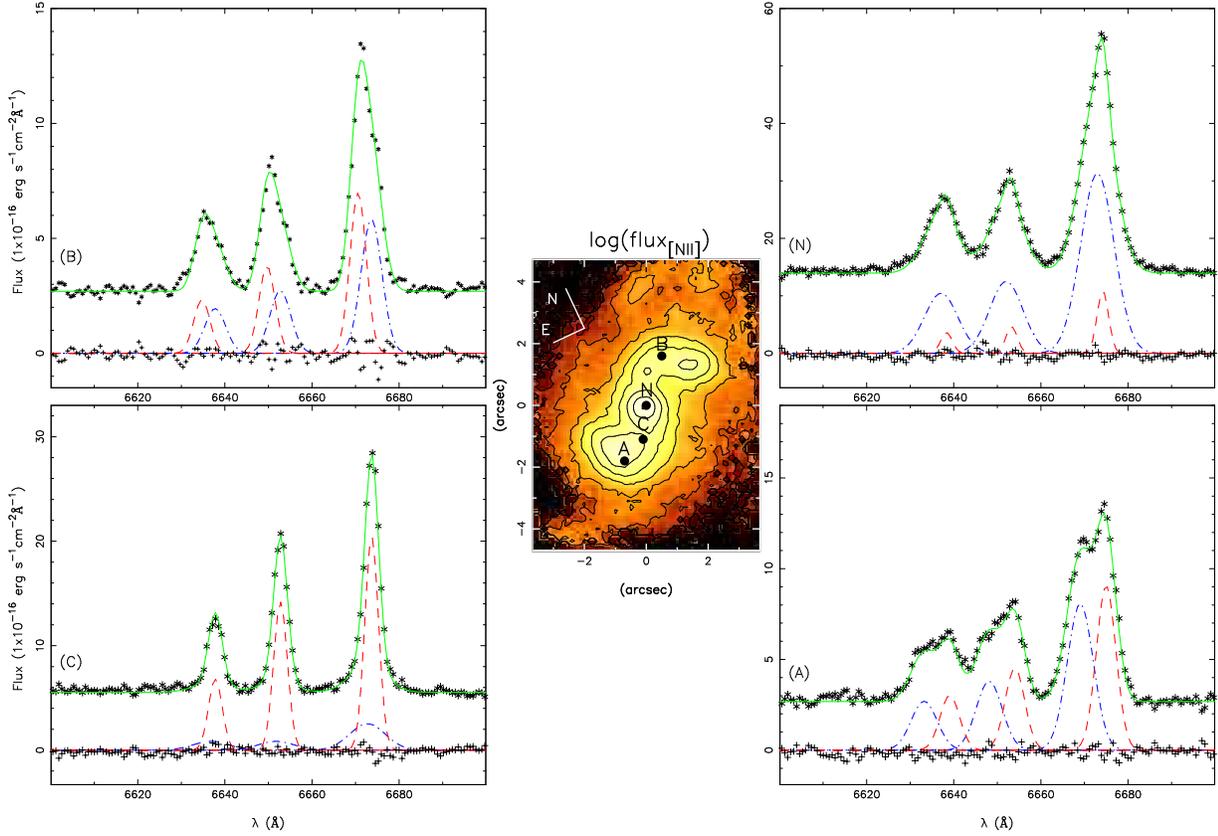}
\caption{Examples of two Gaussian fits to the [N\,II] and H$\alpha$ emission lines from four different regions, labeled as A (x\,=\,-0\farcs7,y\,=\,-1\farcs8), B (x\,=\,0\farcs5,y\,=\,1\farcs6), C (x\,=\,0\arcsec,y\,=\,-1\arcsec) and N (x\,=\,0\arcsec,y\,=\,0\arcsec), selected as representative of the typical double-peaked emission line profiles observed in the inner 2\arcsec. The position of each region is marked on the central box, where the logarithm of the flux of the [N\,II] line is displayed. The asterisks correspond to data points, the solid lines to the fit, crosses to the residuals and the dashed lines to the narrower and broader components. }
\label{figgauss}
\end{figure*}

The gaseous centroid velocities, velocity dispersions and the emission-line fluxes were obtained through the fit of Gaussians to the [N\,II], H$\alpha$, [O\,I] and [S\,II] emission lines. In order to reduce the number of free parameters when fitting the [N\,II] and H$\alpha$ lines, we adopted the following physically motivated constraints: 
\begin{enumerate}
\item Flux$_{[N\,II]\,\lambda6583}$/Flux$_{[N\,II]\,\lambda6548}=2.98$, in accordance with the ratio of
their transition probabilities \citep{osterbrock06};
\item The H$\alpha$, [N\,II]\,$\lambda$6583 and [N\,II]\,$\lambda$6548 lines have the same centroid velocity and FWHM;
\end{enumerate}

As illustrated by the spectra shown in Fig.\,\ref{fig1}, complex emission line profiles which cannot be reproduced by a single Gaussian profile are observed in the nucleus of NGC\,1358. A visual inspection of the datacube showed that double peaked and/or asymmetric line profiles are observed up to $\approx$\,4\arcsec\ from the nucleus. In order to identify distinct kinematic components in the gas, we perform in addition to a single Gaussian fit to the emission line profiles in the entire datacube, a two Gaussian fit to the [N\,II] and H$\alpha$ line profiles in the inner 4\arcsec. If one of the two Gaussian components contributed less than 10\% to the total flux of each emission line in a given spaxel, the two Gaussian fit was discarded. Additionaly, spaxels where the Gaussian parameters (flux, centroid velocity and velocity dispersion) showed large variations compared to neighbouring spaxels were also discarded. Examples of typical two Gaussian fits are shown in Fig.\,\ref{figgauss}. As illustrated by the examples, one of the Gaussian components traces the peak of the line profiles (dashed lines), while the other fits the asymmetries of the line profile (either the broad base, shoulder or the second peak, dot-dashed lines). Although the  [S\,II] line profiles also show asymmetries in the inner 2\arcsec, we could only reliably fit two Gaussian to these lines in few spectra, so we only present single Gaussian fits to [S\,II].

In order to measure the stellar kinematics, we first used the Voronoi Binning technique \citep{voronoi} to bin the datacube in order to achieve a signal-to-noise ratio of at least 5 in the continuum near the Na\,I doublet in each spectrum. We then employed the Penalized Pixel Fitting technique (pPXF, \citealt{cappellari04}), using the \citet{bruzual03} stellar population models as templates to fit the stellar continuum from 5700\,\r{A} to 6600\,\r{A}, to obtain the stellar velocity field and velocity dispersion. 

\subsection{Uncertainties}\label{uncertainties}

\begin{figure*}
\includegraphics[scale=1.1]{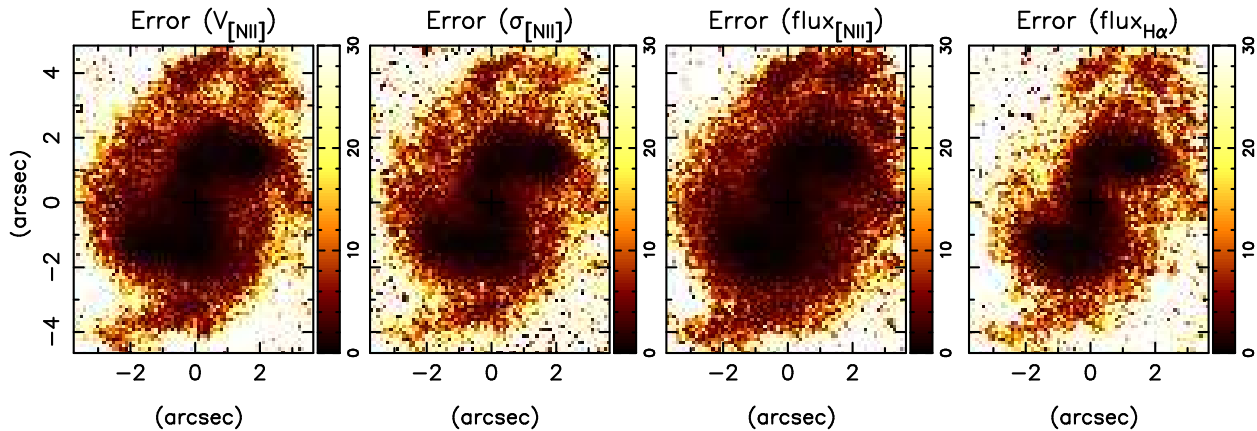}
\caption[Uncertainties]{Uncertainties in centroid velocity (km\,s$^{-1}$), velocity dispersion (km\,s$^{-1}$) for the [N\,II] emission line and uncertainties in flux (\%) for the [N\,II] and H$\alpha$ emission lines. Note that the H$\alpha$ centroid velocity and velocity dispersion were not free parameters in the emission line fits (see section\,3.1 for details).}
 \label{figerror}
 \end{figure*}

To test the robustness of the fits and estimate the uncertainties in the quantities measured from each spectrum in our datacube, we performed Monte Carlo simulations in which Gaussian noise was added to the observed spectrum. For each spaxel, the noise added in each Monte Carlo iteration was randomly drawn from a Gaussian distribution whose standard deviation matches that expected from the noise of that spaxel. One hundred iterations were performed and the estimated uncertainty in each parameter - line center, line width, and total flux in the line - was derived from the $\sigma$ of the parameter distributions yielded by the iterations. In Fig.\,\ref{figerror} we show the uncertainties in the measurement of the [N\,II] emission lines and in the flux distribution of the H$\alpha$ line. Uncertainties in the fluxes of the [S\,II] lines are similar to those of the [N\,II] line. Uncertainties in the stellar velocity and velocity dispersion are of the order of 20\,km\,s$^{-1}$ and 25\,km\,s$^{-1}$ respectively.

\subsection{Stellar kinematics}

\begin{figure}
\includegraphics[scale=1.1]{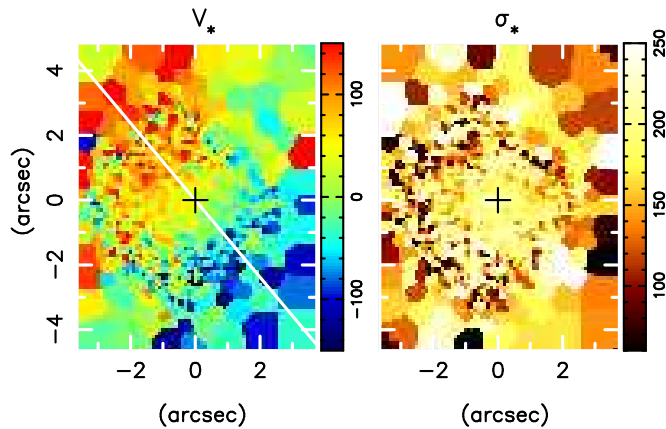}
\caption[Stellar Kinematics]{Stellar centroid velocity (km\,s$^{-1}$) and velocity dispersion (km\,s$^{-1}$), obtained fitting stellar population models to the stellar continuum between 5700\,\r{A} and 6600\,\r{A}. The solid white line marks the position angle of the major axis of the galaxy (15\ensuremath{^\circ}).}
\label{figstars}
\end{figure}

In Fig.\,\ref{figstars} we show the stellar centroid velocity (km\,s$^{-1}$) and velocity dispersion (km\,s$^{-1}$), obtained fitting stellar population models to the stellar continuum between 5700\,\r{A} and 6600\,\r{A}. The stellar velocity field  displays a rotation pattern in which the SW side of the galaxy is approaching and the NE side is receding. Under the assumption that the spiral arms are trailing, it can be concluded that the near side of the galaxy is to the W, and the far side is to the E. The stellar velocity field is consistent with a line of nodes oriented along PA\,=\,15\ensuremath{^\circ}. A systemic velocity of 4029\,km\,s$^{-1}$ (see section.\,\ref{Discussion} for details on how this value was determined) was subtracted from the centroid velocity maps. The stellar velocity dispersion (right panel of Fig.\,\ref{figstars}) varies between 100\,km\,s$^{-1}$ and 250\,km\,s$^{-1}$. 

\subsection{Gaseous Kinematics}\label{kinematics}

\subsubsection{Single Gaussian fit}

\begin{figure*}
\includegraphics[scale=1.2]{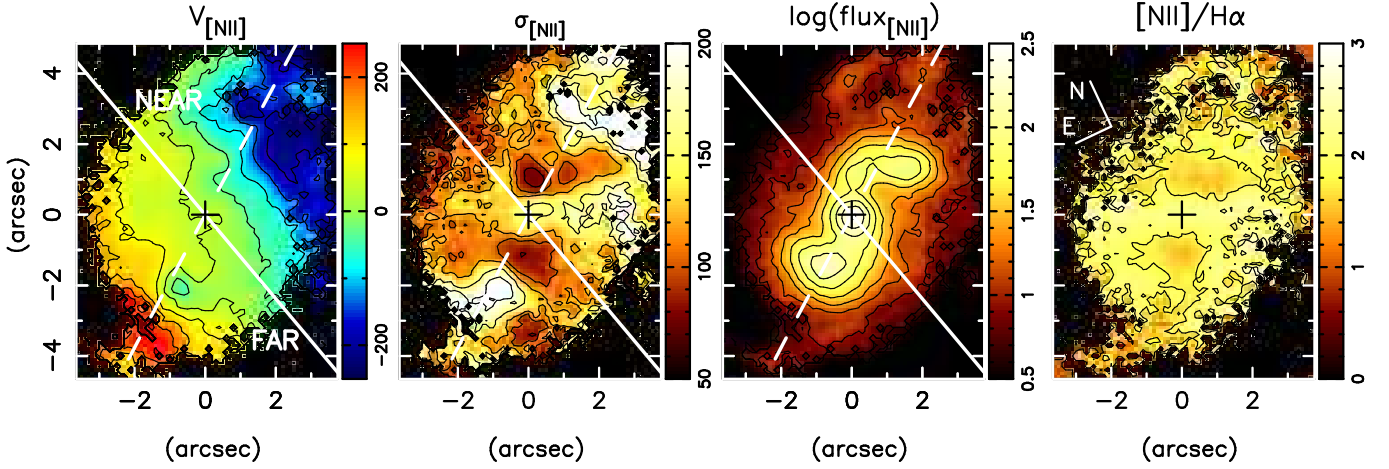}
\caption[Gaseous Kinematics]{Single Gaussian fit: gaseous centroid velocities (km\,s$^{-1}$),  velocity dispersion, logarithm of the [N\,II] emission line flux (in units of 10$^{-17}$\,erg\,cm$^{-2}$\,s$^{-1}$ per spaxel) and [N\,II]/H$\alpha$ ratio. The solid white line marks the position angle of the major axis of the galaxy (15\ensuremath{^\circ}) and the dashed white line marks the position of the large scale bar (135\ensuremath{^\circ}). }
\label{figsingle}
\end{figure*}

In Fig.\,\ref{figsingle} we show centroid velocity (km\,s$^{-1}$),  velocity dispersion, flux distribution, and [N\,II]/H$\alpha$ ratio maps obtained from the single Gaussian fit to the [N\,II] and H$\alpha$ emission lines. A systemic velocity of 4029\,km\,s$^{-1}$ was subtracted from the centroid velocity maps. The gas velocity field is highly disturbed, as evidenced by the strong radial motions near the minor axis of the disk. The velocity dispersion map shows the lowest values (60--80\,km\,s$^{-1}$) to the north-northeast and east-southeast of the nucleus, and the highest values ($\approx$\,200\,km\,s$^{-1}$) between 2--4\arcsec from the nucleus, near the minor axis. High velocity dispersions are also observed around the nucleus in the inner 1\arcsec and along the northeast-southwest direction.

\subsubsection{Two Gaussians fit}

\begin{figure*}
\includegraphics[scale=1.2]{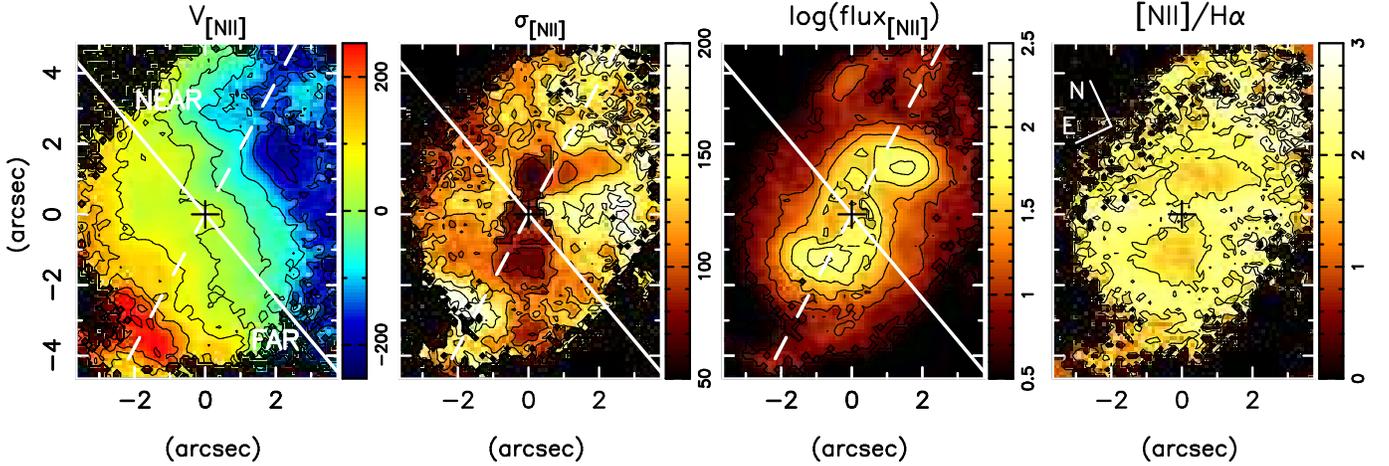}
\caption[Gaseous Kinematics]{Disk component: gaseous centroid velocities (km\,s$^{-1}$),  velocity dispersion, logarithm of the [N\,II] emission line flux (in units of 10$^{-17}$\,erg\,cm$^{-2}$\,s$^{-1}$ per spaxel), and [N\,II]/H$\alpha$ ratio. This map was constructed combining the narrower component and the single Gaussian maps (where only one component was fitted). The solid white line marks the position angle of the major axis of the galaxy (15\ensuremath{^\circ}) and the dashed white line marks the position of the large scale bar (135\ensuremath{^\circ}).}
\label{figdisk}
\end{figure*}

\begin{figure*}
\includegraphics[scale=1.2]{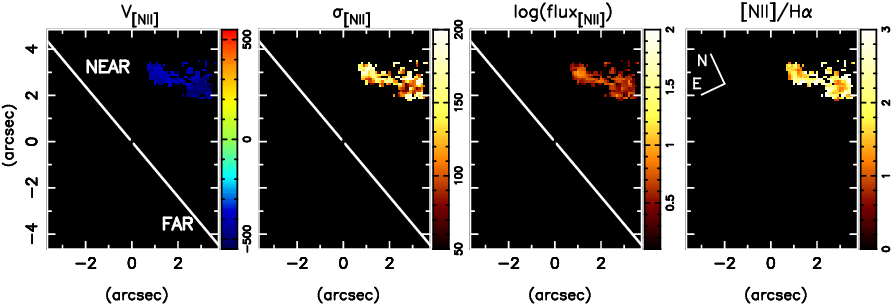}
\caption[Gaseous Kinematics]{Filament component: gaseous centroid velocities (km\,s$^{-1}$),  velocity dispersion, logarithm of the [N\,II] emission line flux (in units of 10$^{-17}$\,erg\,cm$^{-2}$\,s$^{-1}$ per spaxel), and [N\,II]/H$\alpha$ ratio. The solid white line marks the position angle of the major axis of the galaxy (15\ensuremath{^\circ}).}
\label{figfila}
\end{figure*}

\begin{figure*}
\includegraphics[scale=1.2]{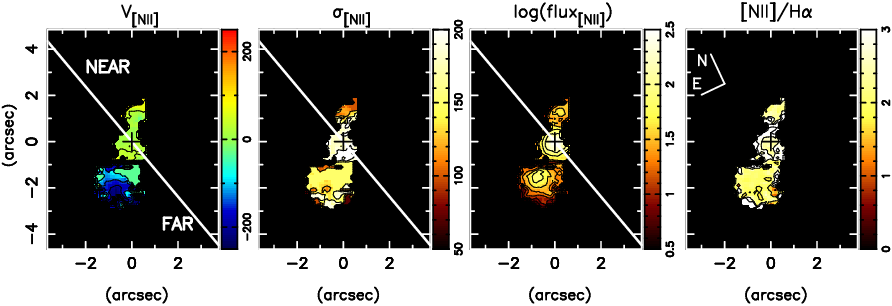}
\caption[Gaseous Kinematics]{Nuclear component: gaseous centroid velocities (km\,s$^{-1}$),  velocity dispersion, logarithm of the [N\,II] emission line flux (in units of 10$^{-17}$\,erg\,cm$^{-2}$\,s$^{-1}$ per spaxel), and [N\,II]/H$\alpha$ ratio. The solid white line marks the position angle of the major axis of the galaxy (15\ensuremath{^\circ}).}
\label{fignuc}
\end{figure*}

A comparison between the [N\,II]  velocity fields obtained from the single and two Gaussian fits showed that the velocity field of one of the components was always consistent with the single Gaussian velocity field. We readily identify this component as due to gas rotating in the disk of the galaxy, and we will hereafter refer to it as the ``disk component". In Fig.\,\ref{figdisk} we show centroid velocity (km\,s$^{-1}$),  velocity dispersion, flux distribution, and [N\,II]/H$\alpha$ ratio maps of the disk component. The largest differences between the disk component and single Gaussian fit maps are observed in the inner 1\arcsec, where the disk component velocities are $\approx$\,40\,km\,s$^{-1}$ larger and in a region 2\arcsec\ southeast of the nucleus, where a blueshifted region is observed in the single Gaussian velocity field while it is not observed in the disk component velocity field (velocities are $\approx$\,100\,km\,s$^{-1}$ larger). The disk component velocity dispersion map shows that there is a large low velocity dispersion region extending from 2\arcsec\ south-southeast of the nucleus to 2\arcsec\ north-northwest. 

We identify two more kinematic components in the two Gaussian fits. One of these components is a extended structure 3--4\arcsec\ northwest of the nucleus, blueshifted by more than 400\,km\,s$^{-1}$ in relation to the systemic velocity. We refer to this component as the ``filament component''. The centroid velocity (km\,s$^{-1}$),  velocity dispersion, flux distribution, and [N\,II]/H$\alpha$ ratio maps of this component are shown in Fig.\,\ref{figfila}. 

The other kinematic component we label ``nuclear component", as it is contained within the inner 2\farcs5. In Fig.\,\ref{fignuc} we show centroid velocity (km\,s$^{-1}$),  velocity dispersion, flux distribution, and [N\,II]/H$\alpha$ ratio maps of the nuclear component. The nuclear component is observed in an elongated region extending from 2\farcs5 south-southeast of the nucleus to 2\arcsec\ north-northwest. The centroid velocity shows large variation across this region. Large blueshifted velocities are observed south-southeast of the nucleus while velocities close to systemic are observed elsewhere. In the velocity dispersion map, three distinct regions are present. A region of high velocity dispersion (180--200\,km\,s$^{-1}$) in the inner 1\arcsec, a region of velocity dispersions of $\approx$\,100\,km\,s$^{-1}$ north-northwest of the nucleus, and a region with velocity dispersions of $\approx$\,150\,km\,s$^{-1}$ south-southeast of the nucleus.

\subsection{Line fluxes and excitation of the emitting gas}\label{fluxgas}

\begin{figure}
\begin{center}
\includegraphics[scale=1.2]{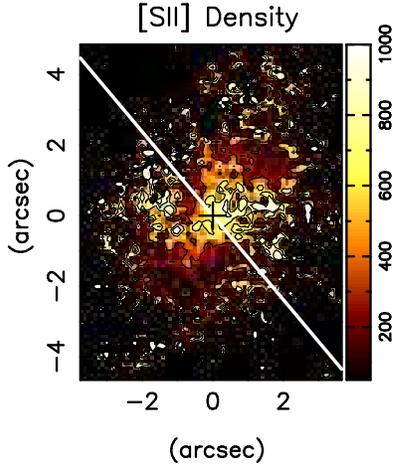}
\caption[Line ratio maps]{Gas density (cm$^{-3}$) obtained from the single component fit. The solid white line marks the position angle of the major axis of the galaxy (15\ensuremath{^\circ}).}
\label{figraz}
\end{center}
\end{figure}

We show the [N\,II] flux distribution for the single Gaussian fit in the center right panel of Fig.\,\ref{figsingle}. The [N\,II] flux distribution for the disk and nuclear components are shown in the center right panels of Fig.\,\ref{figdisk} and Fig.\,\ref{fignuc} respectively. The single Gaussian flux distribution shows an ``S'' shaped structure in the inner 2\arcsec. Three bright emission knots are observed inside this structure. The lowest [N\,II]/H$\alpha$ ratios are observed at the top and bottom of the S-shaped structure. The disk component flux distribution also shows an ``S'' shaped structure, although with a lower flux in the inner 1\arcsec. The nuclear component [N\,II] flux distribution shows three emission knots, one at the nucleus, one in a region 1\arcsec southwest and a fainter knot at 1\arcsec\ north of the nucleus. The [N\,II]/H$\alpha$ ratio varies between 2--3.

In Fig.\ref{figraz} we present the gas density map for the single Gaussian fit. The gas density was obtained from the [SII]\,$\lambda\lambda$6717/6731\,\r{A} line ratio using the \textsc{iraf} task \textsc{temden}, assuming an electronic temperature of 10000K (see Fig.\,5.8 in \citealt{osterbrock06} for a plot of the calculated variation of the line ratio as a function of density for a constant electronic temperature of 10000\,K). 

\section{discussion}\label{Discussion}

\subsection{Stellar Kinematics}

\begin{figure*}
\includegraphics[scale=1.2]{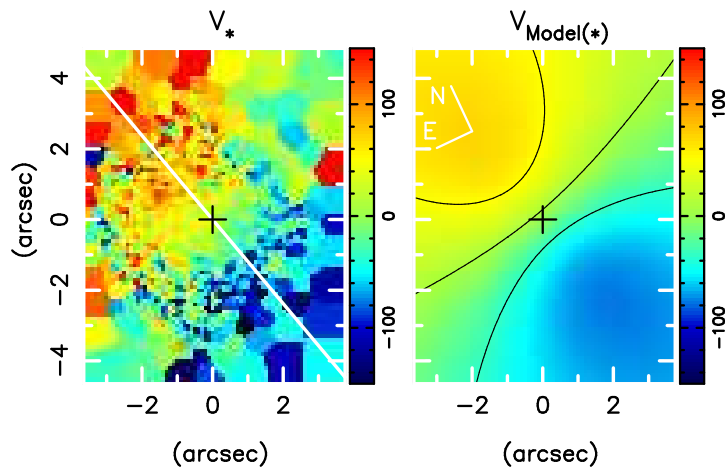}
\caption[Stellar Kinematics]{Stellar velocity field (km\,s$^{-1}$) and modeled velocity field (km\,s$^{-1}$). The solid white line indicates the position angle of the major axis of the galaxy (15\ensuremath{^\circ}).}
\label{figmodel_stars}
\end{figure*}

In order to obtain the value of the systemic velocity and the rotation velocity field, we modeled the stellar velocity field assuming a spherical potential with pure circular motions, with the observed radial velocity at a position ($R,\psi$) in the plane of the sky given by \citep{bertola91}:
\footnotesize
\begin{displaymath}
V=V_{s}+\frac{ARcos(\psi-\theta)sin(i)cos^{p}(i)}{\{R^{2}[sin^{2}(\psi-\theta)+cos^{2}(i) cos^{2}(\psi-\theta)]+c^{2}cos^{2}(i)\}^{p/2}}
\end{displaymath}
\normalsize
where $i$ is the inclination of the disk (with $i$\,=\,0 for a face-on disk), $\theta$ is the position angle of the line of nodes, $V_{s}$ is the systemic velocity, $R$ is the radius in the plane of the sky, $A$ is the amplitude of the rotation curve (at large radii), $c$ is a concentration parameter regulating the compactness of the region with a strong velocity gradient and $p$ regulates the inclination of the flat portion of the velocity curve (at the largest radii). We assumed the kinematical center to be cospatial with the peak of the continuum emission. We adopted an inclination of i\,=\,54\ensuremath{^\circ} (see Table\,1), $p$\,=1 for a assymptoptically flat velocity curve, A\,=\,110\,km\,s$^{-1}$ (from the large scale velocity curve, \citealt{gerssen03}) and a position angle of the line of nodes of 15\ensuremath{^\circ} (see Table\,1). A Levenberg-Marquardt least-squares minimisation was performed to determine the best fitting parameters.

The resulting parameters $c$ and $V_{s}$ are $15\arcsec\pm0\farcs4$ and $4029\,\pm$15\,km\,s$^{-1}$ respectively. Our determination of the systemic velocity is in agreement with the previous determination of $4028\,\pm$10\,km\,s$^{-1}$ based on H\,I 21\,cm measurements (see Table\,1). The model velocity field is shown in Fig.\,\ref{figmodel_stars}.
\subsection{The Disk Component}\label{disk_comp}

\subsubsection{Gas kinematics}
\begin{figure*}
\includegraphics[scale=1.2]{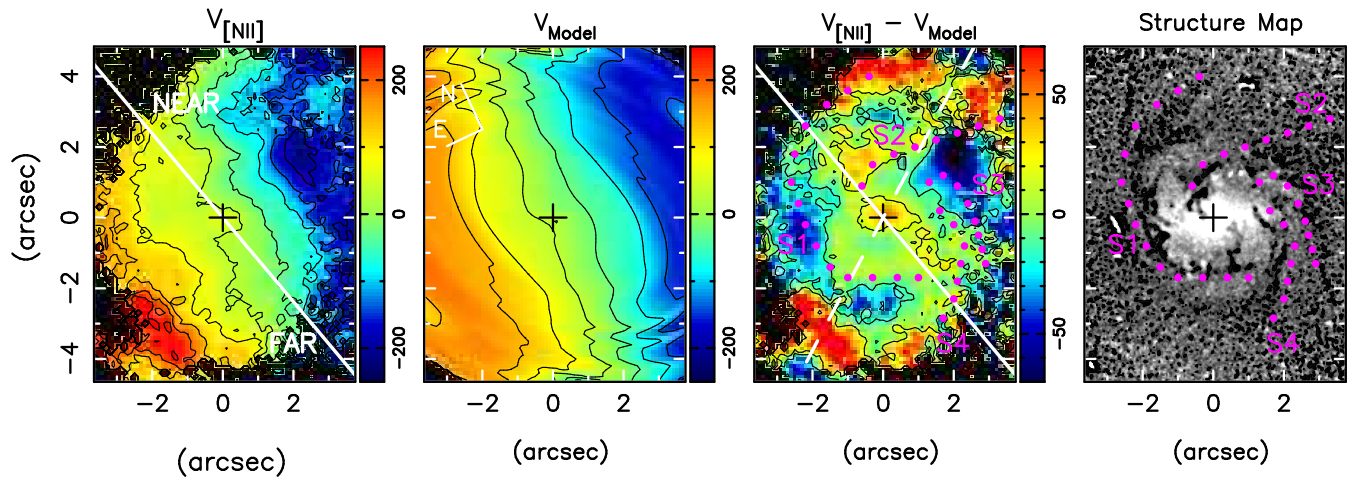}
\caption[Gaseous Kinematics]{From left to right: gaseous centroid velocity (km\,s$^{-1}$), structure map, modelled velocity field and residual between gaseous centroid velocity and modelled velocity field (km\,s$^{-1}$). The solid white line marks the position angle of the major axis of the galaxy (15\ensuremath{^\circ}) and the dashed white line marks the position of the large scale bar (135\ensuremath{^\circ}).}
\label{figmodel}
\end{figure*}

In order to test our hypothesis that nuclear dust structures trace the channels responsible for bringing gas from larger scales to the inner few hundred parsecs, we need to search for radial inflows in the gas. Usually, this is done by fitting a rotating disk model and subtracting it from the observed velocity field. However, in the case of NGC\,1358, the gaseous velocity in the inner few kiloparsecs shows a strong perturbation due to the bar \citep{dumas07} and gas motions in closed orbits cannot be described accurately by a simple rotating disk model. To account for the bar perturbation, we model the gaseous velocity field using the Diskfit code \citep{spekkens07}. Diskfit approximates the observed velocity field in a given position to be:

\begin{equation}
\begin{split}
&V=V_{s}+sin(i)[V_{t}cos(\theta)-V_{2,t}cos(2\theta_b)cos(\theta)\\
&-V_{2,r}sin(2\theta_b)sin(\theta)]
\end{split}
\end{equation}
\normalsize
where $i$ is the inclination of the disk, $V_s$ is the systemic velocity, $V_t$ is the rotation velocity, $\theta$ is the position angle of the disk major axis, $V_{2,t}$ and $V_{2,r}$ are the tangential and radial components of the non-circular bar flow respectively and $\theta_b$ is the angle between the major axis of the bar and the major axis of the disk. As our gaseous velocity field covers only the inner $\approx$\,1\,kpc, which is too small a region to adequately constrain the position angle of the major axis of the disk and bar, we kept these parameters fixed as equal to the corresponding photometric values during the fit. We also fixed the center of the disk as equal to the position of the peak flux in the continuum. 

As our data provides only limited coverage of the inner disturbance and no information on the larger scale undisturbed velocity field, we performed a simple test to assure the velocity field resulting from the fit is meaningful. We combined the observed gaseous velocity field with a large scale velocity field model covering radii of 30$\arcsec$ to 60$\arcsec$ in the plane of the galaxy (where the gaseous velocity field is undisturbed by the bar) and we fitted this combined velocity field with Diskfit. Note that radii not covered by the GMOS field or the large scale model were masked out from the fit. The large scale velocity field was built based on the observed stellar velocity field (which is dominated by rotation). The model and residuals maps were identical to those obtained fitting only the gaseous velocity field. 

We show the gaseous velocity field, model velocity field, residual velocity map and structure map in Fig.\,\ref{figmodel}. White dots tracing the spiral pattern in the structure map are plotted on the residual map. We identify four nuclear spirals in the structure map, which we label S1, S2, S3 and S4. Most of the residuals follow a similar pattern to the dusty spirals, although the kinematical spirals in the residual map are adjacent to the morphological spirals instead of being cospatial. A similar displacement between kinematic and dust spirals has been observed in the region surrounding the AGN in the nearby galaxy NGC\,1097 \citep{fathi06,davies09}. Simulations have also predicted such displacement \citep{maciejewski04a,maciejewski04b}. 

On the near side of the galaxy, redshifted residuals are observed associated to the S1 and S2 nuclear spirals, while on the far side, blueshifted residuals are observed associated to S1. Assuming the gas is on the plane of the galaxy, this means the gas is radially inflowing in these locations. Only one kinematic spiral arm is observed associated to spiral arms S3 and S4, in which redshifted residuals are observed on the far side of the galaxy and blueshifted residuals are observed on the near side, implying that, if the gas is on the plane, it is radially outflowing. In the inner 0\farcs6, redshifted velocities of up to 50\,km\,s$^{-1}$ are observed. This is consistent with measurement of neutral Na\,I gas by \citet{krug10} which found positive velocities of 35$\pm$6\,km\,s$^{-1}$, implying both ionized and neutral gas are inflowing towards the nucleus of NGC\,1358.

A pattern of two inflowing and one outflowing kinematic spirals has previously been observed in the nucleus of NGC\,1097 \citep{fathi06,davies09}. This pattern has been interpreted by \citet{davies09} as a density wave in the disk, associated with a shock, and driven either by the large scale bar or another rotating non-axisymmetric perturbation in the total gravitational potential. In hydrodynamical simulations (see \citealt{maciejewski04b}), this flow pattern emerges as the gas inflowing in the arm  preserves  some  angular momentum, passing by the galaxy center at a certain distance, and continuing as a diverging outflow with smaller gas density \citep{davies09}.

\subsubsection{Line fluxes and excitation}

A remarkable feature in the [N\,II] flux distribution map is the ``S-Shaped'' structure observed in the inner 2\arcsec. Similar structures in the ionized gas emission have been observed in other AGNs and they have been interpreted as due to interaction of the radio jet with gas in the host galaxy (e.g. NGC\,3393, \citealt{maksym16}) or due to illumination of gas in the host by the AGN (e.g. NGC\,2110, \citealt{allan14a}; NGC\,1386, \citealt{lena15}; Mrk\,573, \citealt{fischer16}). The gas velocity dispersion along the ``S'' is somewhat low, varying between 60--120\,km\,s$^{-1}$, and there is no signature of an AGN driven outflow in the residual map, so an interaction of gas disk with a radio jet is unlikely to originate the S-Shaped structure. This leaves illumination of gas in the host by the AGN photoionization cone as the likely origin of this structure. The low velocity dispersion region observed extending along the southeast-northwest direction actually points to this being the orientation of the AGN ionization cone: illumination of kinematically cold gas in the galaxy disk by the AGN can explain the low velocity dispersion. The [O\,III] emission in the inner 5\arcsec\ (see Fig.\,24 in \citealt{mulchaey96}) is also oriented along the southeast-northwest direction, supporting an orientation of the AGN ionization cone along this direction. A comparison between the structure map and the disk component flux distribution shows the tips of the ``S'' are cospatial to the S1 and S3 nuclear spiral arms. Thus, we conclude the S-Shaped structure in the ionized gas flux distribution maps is due to illumination of gas in the disk and in nuclear spiral arms by the AGN. 

\subsection{The Nuclear Component}\label{broader}

From the kinematics and excitation maps in Fig.\,\ref{fignuc}, it is not clear if the nuclear component is composed of a single or multiple structures, as it presents an unusual velocity field and three emission knots are observed. Comparing the nuclear component velocity field to the model velocity field, it is clear the observed velocities are not consistent with rotation in the disk, except for the velocities in the inner 1\arcsec. The nuclear component is observed along the AGN ionization cone, so it could be associated to an AGN driven outflow. In fact, HST long-slit spectra of the inner 1\arcsec\ of NGC\,1358 obtained with the Space Telescope Imaging Spectrograph (STIS) oriented along PA\,=\,24\ensuremath{^\circ} show the ionized gas (H$\alpha$ and [O\,III]) reaches velocities of $\approx$\,100\,km\,s$^{-1}$ and velocity dispersions of $\approx$\,200km\,s$^{-1}$. Redshifted velocities are observed on the far side of the galaxy and blueshifted velocities are observed on the near side, consistent with an outflow. This is in disagreement with our observations, as in the inner 1\arcsec\ the nuclear component velocities are $\approx$\,0km\,s$^{-1}$, and the disk components has low velocity dispersion and redshifted radial velocities on the near side of the galaxy, implying gas inflows not outflows. However, the differences between the HST and GMOS observations can be understood if the outflow observed in the HST data is unresolved in the GMOS observations. This does indeed seem to be the case, as velocities drop to $\approx$\,0km\,s$^{-1}$ at 0\farcs3 from the nucleus in the HST--STIS data (our spatial resolution is 0\farcs6). Thus, we argue the central emission knot in the nuclear component is due to a compact nuclear outflow which is unresolved in our observations. 

Regarding the southeastern and northern emission knots, considering they are observed along the AGN ionization cone, and emission from gas in the disk is observed cospatially to these knots, we suggest they are due to off-plane clouds illuminated by the AGN. These clouds were likely ejected from the nucleus in a previous AGN burst. The difference in the flux distribution of these knots can be understood in this context. The southeastern knot appears brighter as it is in front of the disk, while the northern knot is behind. 


\subsection{Estimating the emitting gas mass}

We can estimate the emitting gas mass in the compact outflow and the clouds from \citep{peterson97}:

\begin{equation}
M\approx2.3\times10^5\frac{L_{41}(H\alpha)}{N_{3}^{2}}M_{\odot}
\end{equation}
 where $L_{41}(H\alpha)$ is the H$\alpha$ luminosity in units of 10$^{41}$\,erg\,s$^{-1}$ and $N_{3}$ is the gas density in units of 10$^3$\,cm\,$^{-3}$. We obtain a mass of emitting gas of 16\,$\times\,$10$^{4}$\,$M_{\odot}$ in the compact outflow (inner 0\farcs8), 64\,$\times\,$10$^{4}$\,$M_{\odot}$ in the southeastern cloud and 5\,$\times\,$10$^{4}$\,$M_{\odot}$ in the northtern cloud.


\subsection{The Filament Component} 

Considering the filament component is blueshifted by more than 400\,km\,s$^{-1}$ in relation to the systemic velocity, this component is likely due to emission from a high latitude gas cloud, photoionized by the AGN. The [N\,II]/H$\alpha$ ratio varies between 1.2--2, similar values to what is observed in the disk component to the along the AGN ionization cone (oriented along the northwest--southeast), consistent with AGN photoionization.   

\subsection{Estimating the mass inflow rate}

\label{inflow}

In the residual velocity map shown in Fig.\,\ref{figmodel}, the gas within $\approx$\,0\farcs7 from the nucleus is observed in redshift. Assuming that this is gas inflowing towards the center, we now calculate the mass inflow rate as: 

\begin{equation}
\dot{M}_{in}\,=\,N_{e}\,v\,\pi\,r^{2}\,m_{p}\,f
\end{equation}
where $N_{e}$ is the electron density, $v$ is the inflowing velocity of the gas , $m_{p}$ is the mass of the proton, $\pi$$r^{2}$ is the are which through which the gas is flowing and $f$ is the filling factor. The filling factor can be estimated from: 
\begin{equation}
L_{H\alpha}\,\sim\,f\,N_{e}^{2}\,J_{H\alpha}(T)\,V
\end{equation}
where $J_{H\alpha}(T)$\,=\,3.534$\,\times\,10^{-25}$\,erg\,cm$^{-3}$\,s$^{-1}$ \citep{osterbrock06} and $L_{H\alpha}$ is the H$\alpha$ luminosity emitted by a volume $V$. Assuming the volume of the inflowing gas region can be approximated by the volume of a cylinder with radius $r$ and height $h$ (distance to the nucleus), we obtain: 
\begin{equation}
\dot{M}_{in}\,=\,\frac{m_{p}\,v\,L_{H\alpha}}{J_{H\alpha}(T)\,N_{e}\,h}
\end{equation}

In the inner 0\farcs7 ($h$\,=\,180\,pc), the average inflow velocity corrected by the inclination of the galaxy is 25\,km\,s$^{-1}$, the average density is 570\,cm$^{-3}$ and the total H$\alpha$ flux is $3.9\,\times\,$10$^{-14}$\,erg\,cm$^{-2}$\,s$^{-1}$. Adopting a distance of 53.7\,Mpc, we obtain $L_{H\alpha}$\,=\,$1.3\,\times\,$10$^{40}$\,erg\,s$^{-1}$. The mass inflow rate of ionized gas in the inner 0\farcs7 is $\dot{M}_{in}$\,$\approx$\,1.5\,$\times\,10^{-2}$M$_{\odot}$\,yr$^{-1}$.

We now compare the estimated inflow rate of ionised gas to the mass accretion rate necessary to produce the luminosity of the Seyfert nucleus of NGC\,1358, calculated as follows:
\[
\dot{m}\,=\,\frac{L_{bol}}{c^{2}\eta} 
\]
where $\eta$ is the efficiency of conversion of the rest mass energy of the accreted material into radiation. For geometrically thin and optically thick accretion disk , the case of Seyfert galaxies, $\eta$\,$\approx$\,$0.1$ \citep{frank02}. The nuclear luminosity can be estimated from the [O\,III] luminosity of $L_{[OIII]}$\,=\,6.0\,$\times\,$10$^{40}$\,erg\,s$^{-1}$ \citep{gu02}, using the approximation that the bolometric luminosity is $L_{Bol}$\,$\approx$\,87$L_{[OIII]}$ \citep{lamastra09}. We use these values to derive an accretion rate of $\dot{m}$\,=\,0.9$\,\times\,$10$^{-4}$\,M$_{\odot}$\,yr$^{-1}$. Comparing the accretion rate $\dot{m}$ with the mass inflow rate of ionised gas, we find that the inflow rate in the inner $\approx$\,180\,pc is about 160 times larger than the accretion rate. We point out, however, that this inflow rate corresponds only to ionised gas, which is probably only a fraction of a more massive inflow in neutral and molecular gas. 

The two orders of magnitude difference between the mass inflow rate in the inner $\approx$\,180\,pc and the accretion rate suggests most of the gas will not reach the nucleus, instead it will accumulate in the inner hundred parsec, building a reservoir which can fuel the formation of new stars. This scenario is supported by the observation of low stellar velocity dispersion regions \citep{emsellem08,comeron08} associated to young to intermediate age (10$^{6}$--10$^{8}$\,yrs old) stellar population \citep{riffel10,riffel11,thaisa12,hicks13} in the inner $\approx$\,200\,pc of nearby Seyferts. 

\section{Summary and Conclusions}\label{Conclusion}

We have measured the stellar and gaseous kinematics of the inner 1.8\,$\times$\,2.5\,kpc$^2$ of the Seyfert\,2 galaxy NGC\,1358, from optical spectra obtained with the GMOS integral field spectrograph on the Gemini South telescope at a spatial resolution of $\approx$\,165\,pc. The main results of this paper are:

\begin{itemize}

\item The stellar velocity field shows rotation in a disk consistent with an orientation for the line of nodes of $\approx$\,15$^\circ$;
 
\item Extended gas emission is observed over the whole FOV, with the line profiles being well fitted by Gaussian curves; 

\item In the inner $\approx$\,650\,pc, four gaseous kinematical components are observed: a component originating in gas rotating in the disk of the galaxy, present over the entire FOV, an unresolved outflow at the nucleus and two off-plane gas clouds, at projected distances of $\approx$\,500\,pc to the southeast and northeast of the nucleus;

\item A fifth kinematical component is observed at $\approx$\,750\,pc north of the nucleus, blueshifted 400\,km\,s$^{-1}$ in relation to the systemic velocity of the galaxy. We interpret this component as a high latitude gas filament;

\item Considering the gas clouds are observed along the AGN ionization cone, we suggest they are due to a previous ejection of the AGN;

\item We estimate an ionized gas mass of $M$\,$\approx$\,16\,$\times\,$10$^{4}$\,M$_{\odot}$ in the compact outflow and $M$\,$\approx$\,64\,$\times\,$10$^{4}$\,M$_{\odot}$ and $M$\,$\approx$\,5\,$\times\,$10$^{4}$\,M$_{\odot}$ in the southeastern and northern gas clouds respectively. 
 
\item The disk component velocity field is strongly disturbed by the large-scale bar. The subtraction of a model combining rotation in a disk and bar flows reveals a three spiral pattern. Residual velocities in these spirals reach up to 80\,km\,s$^{-1}$;

\item We observe residual redshifts associated to spiral arms on the near side of the galaxy and residual blueshifts associated to a spiral arm on the far side. We interpret these residuals as radial inflows;

\item We observe residual redshifts on the far side of the galaxy and blueshifted residuals on the near side associated to another spiral arm. We interpret this residuals as a radial outflow;

\item We have observed a residual redshift within 0\farcs7 of the nucleus, interpreted as due to gas inflow. We have calculated the mass inflow rate in this inflow obtaining  $\dot{M}_{in}$\,$\approx$\,1.5\,$\times\,10^{-2}$M$_{\odot}$\,yr$^{-1}$. This is about 160 times larger than the necessary to power the AGN.
 
\end{itemize}

\section*{ACKNOWLEDGMENTS}

We thank the anonymous referee for comments and suggestions which have improved this paper. This work is based on observations obtained at the Gemini Observatory, which is operated by the Association of Universities for Research in Astronomy, Inc., under a cooperative agreement with the NSF on behalf of the Gemini partnership: the National Science Foundation (United States), the Science and Technology Facilities Council (United Kingdom), the National Research Council (Canada), CONICYT (Chile), the Australian Research Council (Australia), Minist\'erio da Ci\^encia e Tecnologia (Brazil) and south-eastCYT (Argentina). This work has been partially supported by the Brazilian institution CNPq.

\bibliographystyle{mn2e.bst}
\bibliography{ngc1358.bib}

\label{lastpage}
\end{document}